\documentclass[12pt,preprint]{aastex}

\input epsf.sty

\newcommand{\eg}{e.g.\ }

\newcommand{\Msun}{M_{\odot}}

\newcommand{\kms}{km~s$^{-1}$}

\newcommand{\Ha}{H$\alpha$}
\newcommand{\Hb}{H$\beta$}

\newcommand{\HI}{H~{\sc i}}
\newcommand{\HeI}{He~{\sc i}}
\newcommand{\OI}{O~{\sc i}}

\newcommand{\NaI}{Na~{\sc i}}
\newcommand{\MgII}{Mg~{\sc ii}}
\newcommand{\MgI}{Mg~{\sc i}}

\newcommand{\SiII}{Si~{\sc ii}}

\newcommand{\CaII}{Ca~{\sc ii}}

\newcommand{\FeII}{Fe~{\sc ii}}

\newcommand{\CoII}{Co~{\sc ii}}

\newcommand{\Fefs}{$^{56}$Fe}
\newcommand{\Cofs}{$^{56}$Co}
\newcommand{\Nifs}{$^{56}$Ni}
\newcommand{\Mej}{$M_{\rm ej}$}
\newcommand{\KE}{$E_K$}
\newcommand{\vph}{$v_{ph}$}
\newcommand{\fHe}{f_{{\rm HeI}}}
\newcommand{\fH}{f_{{\rm H}}}

\def\gsim{\mathrel{\rlap{\lower 4pt \hbox{\hskip 1pt $\sim$}}\raise 1pt \hbox {$>$}}}
\def\lsim{\mathrel{\rlap{\lower 4pt \hbox{\hskip 1pt $\sim$}}\raise 1pt \hbox {$<$}}}

\shorttitle{The broad-lined SN IIb 2003bg}
\shortauthors{Mazzali et al.}

\begin{document}

\title{SN\,2003bg: a broad-lined Type IIb Supernova with Hydrogen}

\author{
Paolo A.~Mazzali\altaffilmark{1,2,3,4},
Jinsong~Deng\altaffilmark{5},
Mario Hamuy\altaffilmark{6},
Ken'ichi Nomoto\altaffilmark{7}
}

\altaffiltext{1}{Scuola Normale Superiore, Piazza dei Cavalieri, 7, 
  56126 Pisa, Italy} 
\altaffiltext{2}{National Institute for Astrophysics--OAPd, Vicolo
dell'Osservatorio, 5, 35122 Padova, Italy}
\altaffiltext{3}{Max-Planck-Institut f\"ur Astrophysik,
  Karl-Schwarzschildstr.\ 1, 85748 Garching, Germany}
\altaffiltext{4}{Research Center for the Early Universe, University of Tokyo,
  Bunkyo-ku, Tokyo 113-0033, Japan}
\altaffiltext{5}{National Astronomical Observatories, CAS, 20A Datun Road,
  Chaoyang District, Beijing 100012, China}
\altaffiltext{6}{Departamento de Astronom\'\i a, Universidad de Chile, 
  Casilla 36-D, Santiago, Chile}
\altaffiltext{7}{Institute for the Physics and Mathematics of the Universe,
University of Tokyo, Kashiwa, 5-1-5 Kashiwanoha, Kashiwa, Chiba 277-8568, Japan}

\begin{abstract}
Models for the spectra and the light curve, in the photospheric as well as in 
the late nebular phase, are used to infer the properties of the very 
radio-bright, broad-lined type IIb Supernova 2003bg. Consistent fits to the 
light curve and the spectral evolution are obtained with an explosion that 
ejected $\approx 4 \Msun$ of material with a kinetic energy of 
$\approx 5\:10^{51}$\,erg. A thin layer of hydrogen, comprising 
$\sim 0.05 \Msun$, is inferred to be present in the ejecta at the highest 
velocities ($v \gsim 9000$\,\kms), while a thicker helium layer, comprising 
$\approx 1.25 \Msun$, was ejected at velocities between 6500 and 9000\,\kms. 
At lower velocities, heavier elements are present, including $\sim 0.2 \Msun$ 
of \Nifs\ that shape the light curve and the late-time nebular spectra. These 
values suggest that the progenitor star had a mass of $\approx 20-25 \Msun$
(comparable to, but maybe somewhat smaller than that of the progenitor of the 
XRF/SN 2008D). The rather broad-lined early spectra are the result of the 
presence of a small amount of material ($\approx 0.03 \Msun$) at velocities 
$> 0.1 c$, which carries $\approx 10$\% of the explosion kinetic energy. 
No clear signatures of a highly aspherical explosion are detected.
\end{abstract}


\section{Introduction}

Stars with an initial mass exceeding $8-10 \Msun$ end their life when their core
collapses to a compact remnant and the outer layers are ejected in a powerful
explosion, a Supernova (SN). If the outer hydrogen envelope of the star is still
in place at the time of the explosion, the SN is characterized by strong H lines
and is classified as type II \citep[see][for a primer on SN
classification]{fil97}. There are however cases when the outer layers of the
star have been lost before the explosion. In particular, if the H envelope was
lost but the He envelope was retained, the SN shows He lines and is classified
as a type Ib, while if both the H and He shells were lost neither H nor He lines
are seen and the SN is classified as type Ic. Collectively, the class of
stripped-envelope SNe is called SNe\,Ib/c. Type Ib SNe are rarer than type Ic
SNe.  These subclasses should not be thought of as rigid separations, but rather
as an indication of what is probably a continuum of properties.  For example,
evidence of residual H lines has been claimed in some SNe\,Ib 
\citep[\eg][]{deng00,anup05}.

The study of stripped-envelope SNe gained momentum with the discovery that some
extremely energetic SNe\,Ic (``hypernovae'': HNe) are linked to Gamma-Ray Bursts
\citep{gal98,iwa98}. Thereafter, various studies of SNe\,Ib/c have been
performed, both of individual SNe [\eg SN\,1999ex \citep{ham02}; SN\,2004aw
\citep{taub06}], and systematic \citep[\eg][]{mat03,sod06}. 

Some SNe\,Ib/c have also been found in association with X-ray Flashes (XRF), the
softer and weaker versions of GRBs \citep{heise01}: the SN\,Ic 2006aj
\citep{camp06,pian06} and the SN\,Ib 2008D \citep{sod08}. While the nature of
the XRF is debated in both cases \citep{camp06,ghis07,wax07,sod08,maz08}, both
SNe turned out to be less energetic explosions than the GRB/SNe, but still
significantly more energetic than typical SNe\,Ib/c \citep{maz06,maz08,tan09},
which release $\approx 10^{51}$\,erg of energy \citep[\eg][]{iwa94,sau06}.


It is now clear that SNe\,Ic, and to some extent also SNe\,Ib, show a large
spread of properties \citep[\eg][]{cloc97}. Studies of various individual
objects have begun to paint a picture where the so-called `typical' SNe\,Ib/c,
such as SN\,1994I [\eg \citet{nom93,nom94,wheel94,fil95,rich96}] are just a
subset of this extended family. Thus, the definition of typical may need
revising. The mass of the progenitor stars of SNe\,Ib/c is not constrained,
except that  it must be larger than the minimum mass to reach core collapse ($8
- 10 \Msun$), if the progenitor was a single star. 

The final fate of massive stars depends mostly on mass, but also on other
properties, such as metallicity, rotation, and magnetic fields
\citep[\eg][]{fryer99,heger03,W&J05}. These dependencies on often poorly known
quantities add some uncertainty to the determination of the properties of the
progenitor from those of the SN ejecta. Presently, SN\,Ib/c observations and
models indicate that the basic SN properties (\eg the kinetic energy liberated
in the explosion, \KE, the mass ejected, \Mej) are apparently related to the
progenitor's mass \citep{nom05}. It takes very massive stars (M$_{ZAMS} \sim 40
\Msun$), possibly of low metallicity, to make a GRB in a highly aspherical
explosion that leaves behind a black hole \citep{fryer99,McFW}. Less massive
stars may still collapse to a black hole, possibly via fallback \citep{fryer99},
but not give rise to a GRB [\eg SNe 2002ap, \citet{maz02} or 1997ef,
\citet{iwa00,maz01}]. At lower masses still [M$_{ZAMS} \lsim 22 \Msun$,
\citep{fryer99}, at least if the effect of rotation is not considered], the star
collapses to a neutron star and may eject some material at relativistic
velocities, but in this case the relativistic outflow is less energetic and is
observed as an X-Ray Flash [\eg SN\,2006aj \citep{maz06}]. These less energetic
explosions may be less asymmetric \citep{maz07,mae08}. 

One general feature of SNe Ib/c is that their progenitors must have lost at
least the outer hydrogen envelope. If a hydrogen envelope more massive than
$\sim 0.1 \Msun$ is left, the star would probably have a RSG structure at the
time of core collapse, and the SN light curve would be influenced initially by
hydrogen recombination and appear as a type II light curve. Some transitional
objects like SN\,1993J only had a very small hydrogen envelope, whose presence
did not affect the light curve significantly but was responsible for the Type II
spectrum this SN displayed early on \citep{nom93}. Spectroscopically, after an
early Type II phase SN\,1993J turned to a Type Ib, and is therefore known as an
SN\,IIb. Only few SNe\,IIb have been observed, and little is known about their
spread in mass and energy. 

Exactly how a star manages to preserve just a thin shell of H is debated. The
scenarios are different depending on whether the star was single or a member of
an interacting binary. In the former case, a mass loss occurs via a radiatively
driven stellar wind \citep[\eg][]{weil07,stock07}. This can be effective at
relatively high masses ($\gsim 30 \Msun$). Stars with masses of $\sim 27 - 28
\Msun$ may end their lives with only a thin H shell left in place
\citep{crockett08}. Such stars would be classified as WRL or WRH
\citep{smith&conti08}.  However, this mechanism is much less efficient for less
luminous stars. Yet, the inference for events such as SN\,1993J is that the
progenitor had a mass of $\approx 15 \Msun$. In this case the latter scenario,
binary interaction, seems more appealing. For example, \citet{nom95} suggest
that an SN\,IIb is a result of the spiral-in of a low-mass companion into a
massive star, which causes the ejection of most of the massive star's H-rich
envelope.  This is probably a good scenario to produce an energetic explosion,
because it can explain both how the core gains angular momentum from spiral-in,
and how most of the envelope is ejected by spiral-in heating. There is some
observational evidence in support of the binary scenario: for SN\,1993J a
spectroscopic companion has been observed \citep{maund04}. Evidence of a binary
companion has also been claimed for the SN\,IIb 2001ig \citep{ryder06}, while
for the SN\,Ib 2008ax \citet{crockett08} suggest that binary interaction may be
a more viable scenario than that of a single star. A similar debate exists also
regarding the origin of SNe Ib/c \citep[\eg][]{nom93,nom01}.   

SN\,2003bg was initially classified as a broad-lined SN\,Ic \citep{filcho03},
but it soon developed a broad \Ha\ P-Cygni feature as well as \HeI\ lines, 
making it a SN\,IIb. This is the first SN\,IIb for which broad lines have been
observed, suggesting that it is a SN\,IIb hypernova. The optical data of
SN\,2003bg are presented in a companion paper \citep{ham09}. In the nebular
phase the spectra were similar to those of SNe\,Ib/c, but evidence of H may be
present in the form of a weak, flat-topped \Ha\ emission, reminiscent of
SN\,1993J. Additionally, SN\,2003bg was exceptional for its radio luminosity.
\citet{sod06b} suggest that the observed radio light curve can be interpreted as
the signature of sub-relativistic ejecta (with velocity $v \sim 0.2 c$)
producing synchrotron radiation as they impacted into a medium that may have had
the properties of a WR-star wind (mass loss rate $\dot{M} \sim 4\:10^{-5}
\Msun$\,yr$^{-1}$, wind velocity $v \sim 1000$\,\kms). They also suggest that
the wind may have been highly variable, especially just before the explosion.
This is reminiscent of the situation of the XRF-SN\,2006aj, which was probably
surrounded by a thick wind shell [\citet{camp06,wax07}, but see 
\citet{ghis07}]. Interestingly, the radio light curve of SN\,2003bg and that of
SN\,2001ig were very similar.  

The available evidence therefore suggests some similarity among all SNe\,IIb,
despite clear differences in the ejected mass and the kinetic energy of the
explosion: for SN\,1993J \citet{nom93} estimated an ejected mass of $4 \Msun$
and an explosion \KE\ of $10^{51}$\,erg, while for SN\,2001ig
\citet{silverman09} estimate an ejected mass of $\sim 1 \Msun$ below a velocity
of 4300\,\kms. 

Here we concentrate on extracting the basic properties of SN\,2003bg, and hence
of the progenitor star and of the explosion. In order to do this we model the
light curve and the spectra in both the early photospheric phase and the late
nebular phase.

\section{Data}

Observations of SN\,2003bg started very soon after the putative explosion date
(2003 Feb 25 UT), and cover the evolution of the SN for almost one year,
extending into the nebular phase \citep[see][for more details]{ham09}. The peak
phase was unfortunately not covered, and there is a gap in the data at the onset
of the linear decline phase. Still, the available data make it possible to
constrain the properties of the SN reasonably accurately, especially in view of
the availability of both early and late spectra. 

We have modelled 10 spectra, of which 8 were obtained around the peak and 2 in
the nebular phase. Combining the analysis of the outer part of the ejecta
conducted by means of the early-time spectra and the determination of the
properties of the inner, slow-moving ejecta made possible by the nebular
spectra, both the density and the abundance stratification in the SN ejecta can
be determined.

\section{Models for the early phase}

Unlike Type Ia SNe, the modelling of SNe Ib/c must establish both \KE\ and \Mej\
simultaneously. As these are the two parameters that play the major role in
determining both the light curve and the spectra, clearly a reliable solution
must be able to reproduce simultaneously both the spectral evolution and the
light curve. As is well known, the solution of the light curve alone is
degenerate to \KE\ and \Mej\ if both are allowed to vary \citep{arn82}, as is
the case for SNe\,Ib/c. 

The first steps to obtaining a solution are therefore to consider that: {\rm i})
the light curve has a rather narrow peak: the presence of H and He does not
influence the light curve much, because the H shell must be thin and He does not
contribute significantly to the opacity. {\rm ii}) the earliest spectra have
broad lines, similar to but not quite as broad as SN\,2002ap \citep{maz02}. {\rm
iii}) the brightness of SN\,2003bg at peak suggests that it synthesized $\sim
0.2 \Msun$ of \Nifs. This is more than in `normal' SNe\,Ib/c, and comparable to
non-GRB hypernovae such as SNe\,1997ef or 1997dq \citep{maz04}, and the XRF-SNe
2006aj \citep{maz06} or 2008D \citep{maz08,tan09}.

\subsection{Method}

The early-time spectra were modelled using a Monte Carlo radiative transfer code
\citep{m&l93,l99,maz00}. The code assumes that the SN luminosity emerges from a
sharp, grey photosphere (Schuster-Schwarzschild approximation). This
approximation is valid if most $\gamma$-rays and positrons produced in the
radiactive decay of \Nifs\ to \Cofs\ and hence to \Fefs\ deposit their energy,
which is eventually thermalised and converted into optical photons, below the
photosphere. This is the case in the early phases of a Type I SN. Using as input
the SN luminosity and the velocity of the photosphere at the desired epoch of
the model (velocity and time give the photospheric radius since the expansion of
SN ejecta is homologous: $v = r t$), an approximate effective temperature at the
photosphere can be obtained assuming black body emission ($L = 4 \pi R^2
\sigma_B T^4$, where $\sigma_B$ is Stefan-Boltzmann's constant). 

Photons emitted at the photosphere propagate in the expanding ejecta, where
radiative equilibrium is assumed.  Photons can undergo electron scattering and
they can interact with spectral lines. In the latter case, photon branching is
explicitly considered \citep{l99,maz00}. This process allows photons to shift
from wavelengths where line opacity is very high to spectral regions where line
opacity is low, and thus to escape the SN envelope, and it is essential in
shaping the spectra of Type I SNe.  Level populations and the ionization
structure in the SN envelope are computed  taking into account deviations from
Local Thermodynamic Equilibrium (LTE) using a nebular approximation as discussed
in \citet{abb&l85} and \citet{m&l93}. This approximation has been shown to
reproduce well the effects of non-LTE \citep{paul96}.  Iterating a series of MC
experiments, a temperature structure in the ejecta is then computed. Finally,
the emerging flux is recorded using an efficient formal integral approach
\citep{l99}. The calculation of the observed flux requires the values of the
distance and the reddening to the SN. The code has been applied to a number of
Type I SNe \citep[\eg][]{maz08}. 

Apart from the parameters described above, and for atomic models, the only other
input required by the code is the run of density in the ejecta with expansion
velocity. This is known as an explosion model. The properties of the model
determine the appearance of the spectra. The basic quantities that define an
explosion model are the ejected mass and the kinetic energy of the explosion. In
particular, the presence of large amounts of material at high velocities leads
to spectra characterized by broad lines \citep{min00}, while, given the same
kinetic energy, more massive models have broader light curves \citep{arn82}. The
simultaneous fit of the light curve and the spectra of an SN using one explosion
model therefore constitutes a validation of the model itself.

Because of the spectral similarity with SNe 1997ef and 2002ap we started with an
explosion model based on CO100, a model characterized by \Mej $\sim 8 \Msun$ and
\KE $\sim 10^{52}$\,erg. Model CO100 was used to reproduce the light curve and
the spectral evolution of SN\,1997ef, but the very broad lines observed in this
SN required the addition of  a high-velocity, low-density tail which increased
the \KE\ to $\sim 2\:10^{52}$\,erg. The model was also used in rescaled form
(1/4 of the mass) to fit the SN\,Ic 2002ap, which was characterized by broad
spectral features but a relatively narrow light curve \citep{maz02}. SN\,2003bg 
poses a similar problem. In order to reproduce its narrow light curve we needed
to rescale model CO100 down in mass. Furthermore, line blending in SN\,2003bg is
less extreme than in SNe\,1997ef or 2002ap - it is more similar to SN\,2006aj:
\eg the \CaII\ IR triplet and the \OI\ 7774\,\AA\ line do not blend - and so the
density gradient in the outer part must have been steeper. Therefore we did not
need the high-velocity tail, and we finally used a model with half the mass of
the original CO100. The model we selected has \Mej\,$\sim 4 \Msun$ and
\KE\,$\sim 5\:10^{51}$\,erg. 

In the spectral models the luminosity $L$, which is a required input, was
determined such as to match the flux in the spectra. Another required input is
the photospheric velocity \vph. This was chosen so that the position of the
absorption lines and the overall spectral distribution in the synthetic spectra
reproduced the observations as closely as possible. The abundances in the
ejecta, also a required input, were chosen to optimize the fit, but our aim was
not so much to reproduce the details of the spectra but rather to explain the
general evolution of the SN. 

Because H and He lines appear in the spectra we assumed that these elements are
present in the outer part of the ejecta. He lines are notoriously not visible
unless the levels of \HeI\ are non-thermally excited \citep{l91}. Non-thermal
excitation is caused by the fast electrons which are produced in the
thermalization process of the $\gamma$-rays that are emitted by the radioactive
decay of \Nifs\ into \Cofs\ and hence into \Fefs, which is the source of light
for SNe. At early times, the density in the ejecta is high and thermalization
occurs near the place where the $\gamma$-rays are emitted. Since the helium
layer is located at a larger radius than the \Nifs\, non-thermal processes do
not occur. Only at later times do fast electron begin to travel considerable
distances before being thermalized. In fact, \HeI\ lines develop rather late in
SNe\,Ib \citep[\eg SN\,2008D,][]{maz08}.

We do not explicitly treat the propagation of the fast electrons, which also
requires a detailed mapping of the distribution of \Nifs\ in what may have been
an aspherical explosion. Therefore, the departure coefficients caused by
non-thermal processes with respect to the level population computed with our
pseudo-NLTE treatment are not explicitly computed. Instead, we mimic the effect
of overexcitation and ionization caused by non-thermal processes using an
arbitrary `non-thermal factor' $\fHe$ to multiply all \HeI\ excited level
populations. These factors increase with time \citep{m&l98}, reflecting the
increasing penetration of fast electrons in the SN envelope, and they were taken
to be similar to those derived by \citet{l91}. While this is a rough treatment,
it gives one at least a feeling for the presence or lack of He in the ejecta
\citep{tom05}. A non-thermal factor was also used for hydrogen ($\fH$): since
the H mass must be small, this was required to match the observed line
strengths.

The explosion date was chosen to be 2003 February 25. This gives a rise time to
maximum of $\sim 15 - 20$ days, similar to SN\,1998bw. We used a distance
modulus of 31.68 ($d = 24$\,Mpc), and a very small reddening [$E(B-V) = 0.02$].
The early-time spectral fits are presented in Figures 1 and 2 and are briefly
discussed in the following. The input values for the spectral models are
summarized in Table 1.

\subsection{Results}

\subsubsection{2003 Feb. 28}

The model is shown in Figure 1a. It was computed using $\log L = 41.47$ [erg
s$^{-1}$], \vph$ = 22000$\,\kms, and $t = 3$\,days. The mass above the
photosphere is $\approx 0.2 \Msun$, and the radiation temperature at the
photosphere is $\sim 7000$\,K. Although He dominates the composition (90\% by
mass above the photospheric velocity), and a non-thermal factor $\fHe = 10^3$
was introduced, no strong He lines are visible, and so the He distribution
cannot be constrained. Hydrogen is present only in a very small amount (4\% by
mass). If more H is included, the increased electron density changes the
ionization balance of all other species. Therefore, the same non-thermal
excitation factor as for \HeI\ was used also for \HI\ ($\fH = 10^3$).  Both H
and He lines make only a minor contribution to the troughs near 6000 and
4300\AA, which are dominated by \SiII\ and \CoII\ lines, respectively. Other
features are due to \FeII, \SiII, \OI, \CaII, and \MgII\ and are marked in
Figure 1a. The presence of a small amount of material at high velocity ($\approx
0.03 \Msun$) are located at $v > 0.1 c$ is sufficient to give rise to broad
lines.

\subsubsection{2003 March 2}

The model shown in Figure 1b was computed with $\log L = 41.82$ [erg s$^{-1}$],
\vph$ = 18000$\,\kms, and $t = 5$\,days. It is similar to the previous one, but
the lines are deeper as the photosphere moved rapidly inwards. The mass above
the photosphere is now $\approx 0.4 \Msun$. The abundances are also similar,
with He the dominant species. The higher $L$ leads to a higher radiation
temperature near the photosphere, $T \sim 7500$\,K. Hydrogen lines now make an
important contribution: in the feature near 6200\,\AA, \Ha\ is as strong as
\SiII\ 6355\AA, which contributes to making that feature broad. The same
non-thermal factors as for Feb 28 ($\fHe = 10^3, \fH = 10^3$) were used for both
\HI\ and \HeI.  \HeI\ lines are present in the spectrum. \HeI\ 5686\,\AA\ is
blended with \NaI\ D, which is however weaker. The inconsistent strength of \Ha,
which is too strong, and \Hb, which is comparable to the observation, may
suggest that a full, rather than a parametrized treatment of NLTE and
non-thermal effects for at least H may be required. If the contribution of
\SiII\ 6355\AA\ to the \Ha-\SiII\ feature is reduced, the synthetic line does
not match the velocity of the observed absorption.

\subsubsection{2003 March 4}

This model is shown in Figure 1c. It was computed with $\log L = 41.95$ [erg
s$^{-1}$], \vph$ = 14100$\,\kms, and $t = 7$\,days. It has a similar composition
as the previous model, and the same non-thermal factors for \HI\ and \HeI\
($\fHe = 10^3, \fH = 10^3$). As the temperature continues to increase (now $T
\sim 8000$\,K) with the luminosity, and more mass (now $\approx 0.7 \Msun$) is
located above the photosphere, the H and He lines make an increasingly important
contribution. \Ha\ is now stronger than \SiII\ 6355\,\AA, \Hb\ is visible, and
so are some of the \HeI\ lines (especially 4471\,\AA\ and 5876\,\AA). Apart from
this, there are no major differences with respect to the previous spectrum.

\subsubsection{2003 March 12}

This is the spectrum nearest the time of maximum, the epoch of which was not
covered. It was computed with $\log L = 42.31$ [erg s$^{-1}$], \vph$ =
8800$\,\kms, and $t = 15$\,days, and it is shown in Figure 1d. The spectrum is
now bluer than before, but the lines are the same. The temperature has further
increased ($T \sim 8600$\,K), and now \HI\ and \HeI\ lines dominate (the
composition is similar to the previous epochs, despite the much larger mass
located above the photosphere, $\approx 1.5 \Msun$, and the same non-thermal
factors were used for both \HI\ and \HeI: $\fHe = 10^3, \fH = 10^3$), while Fe
makes a smaller contribution by comparison.

\subsubsection{2003 March 31}

This spectrum was obtained after an observational gap near the time of maximum.
It is much redder than all pre-maximum spectra, which is the result of the
decrease in temperature ($T \sim 6000$\,K) and luminosity. The model was
computed with $\log L = 42.29$ [erg s$^{-1}$], \vph$ = 6000$\,\kms, and $t =
34$\,days, and it is shown in Figure 2a. The mass above the photosphere is now
$\approx 2.4 \Msun$. The abundances are not very different, with a slight
decrease of He, which is reduced to $\sim 70$\% and replaced mostly by Ne near
the photospheric layers. The abundance of Fe-group elements has now increased to
$\sim 4$\%, which is required to block the near-UV flux. 

Several \HeI\ lines are now clearly visible, emphasizing the IIb nature of the
SN. They can be reproduced using $\fHe = 10^6$. An increase with time of the
departure coefficients can be expected because of the increased penetration of
the $\gamma$-rays, as discussed by \citet{m&l98}. Although any statement about
the He mass is made highly uncertain by the presence of the huge non-thermal
correction, the strength of the He lines suggests that the He mass should be
large. In any case, the He-shell is limited to $\sim 2 \Msun$ at most stellar 
masses in evolutionary models \citep{nom88}.  

The \HI\ lines are now narrower, and have a smaller blueshift than lines of
other elements. In order to get a reasonable match to these lines (only \Ha\ and
\Hb\ are seen) it must be assumed that H is present only at $v \gsim
9000$\,\kms. The non-thermal coefficient is only marginally larger than at
previous epochs ($\fH = 2\:10^3$). Given the low abundance of H (less than 4\%),
we derive a total H mass of $\sim 0.05 \Msun$. The mass of H cannot be much
larger for two reasons. Firstly, if the mass was much larger than $0.1 \Msun$
the star would not be compact and one would expect a type II light curve, with a
phase dominated by H recombination. Secondly, a large H mass would imply a
higher electron density, which would then result in a different overall
ionization regime and in a worse fit to the spectra. The oxygen abundance near
the photosphere cannot be determined because the spectra do not cover the \OI\
7774\,\AA\ line, so the same value as in the previous epoch was used (3\%).

\subsubsection{2003 April 4}

This spectrum was obtained just a few days later and is similar to the previous
one. The model (Figure 2b) was computed with $\log L = 42.24$ [erg s$^{-1}$],
\vph$ = 5600$\,\kms, and $t = 38$\,days. The mass above the photosphere is now
$\approx 2.5 \Msun$, and the temperature has decreased only slightly, to
$5600$\,K. The lines are now more clearly separated, with metal lines making a
larger contribution to the spectrum, although \Ha, \Hb, and some He lines are
still very strong. The abundances have not changed much. The non-thermal
coefficient for He is larger ($\fHe = 10^7$),  but comparable to what was used
in \citet{tom05}.  The non-thermal coefficient for H is the same as in the
previous epoch ($\fH = 2\:10^3$). H is present only above 9000\,\kms, but now a
confinement for He has also been introduced at 6500\,\kms. This is required to
fit the shape and position of the \HeI\ lines.

\subsubsection{2003 April 9}

The next spectrum is similar to the previous one. The model (Figure 2c) was
computed with $\log L = 42.21$ [erg s$^{-1}$], \vph$ = 5100$\,\kms, and $t =
43$\,days, and has a mass above the photosphere of $\approx 2.65 \Msun$. The
spectrum is marginally cooler than the previous one ($T \approx 5300$\,K, but
the lines that are present are the same. \HI\ and \HeI\ lines are sharp. The
abundances have changed consibrably with respect to the previous spectrum. The
abundance of He is down to $\sim 50$\% by mass, and O, Ne, and Si are highly
abundant near the photosphere.  The confinements of H (above 9000\,\kms) and He
(above 6500\,\kms) are as in the previous spectrum, but the non-thermal factors
are higher: $\fHe = 10^9$ and $\fH = 3\:10^3$. Departure coefficients that
increase with time are expected as $\gamma$-rays propagate more efficiently
\citep{m&l98}, but the precise values should be tested with realistic
calculations. These require a detailed knowledge of the distribution of \Nifs\
as well as of its mass, and are postponed to future work. Still, a slower rise
of the non-thermal factor for \HI\ than for \HeI\ may be justified by the fact
that hydrogen is located further out than helium in the ejecta, and hence it is
more separated from the source of fast electrons. Also, \HI\ levels are
separated by smaller energies than those of \HeI, and thus are more coupled to
the thermal pool. This also applies to the ionization potentials of the two
ions.

\subsubsection{2003 April 10}

This is the last of the photospheric epoch spectra. The model (Figure 2d) was
computed with $\log L = 42.20$ [erg s$^{-1}$], \vph$ = 5000$\,\kms, and $t =
44$\,days. The composition, non-thermal factor and confinements are the same as
in the previous spectrum, which was only one day earlier. The mass above the
photosphere is now $\approx 2.7 \Msun$, and the temperature is $\approx
5300$\,K. \HI and \HeI\ lines are very sharp. The volume sampled by last few
spectra in the photospheric epoch is just outside the volume sampled by the
nebular spectra.

\subsection{Results of early-phase modelling}

The main result of the photospheric-epoch models are that SN\,2003bg was a very
energetic explosion (a hypernova) and that H and He are both present, but are
clearly confined in velocity: H above 9000\,\kms\ and He above 6500\,\kms. This
explains the weakness of the H and He lines in the early spectra and the almost
complete absence of H in the nebular spectra: these elements are removed from
\Nifs, the source of the fast particles responsible for non-thermal processes. 
Thus the SN looked like an SN\,Ic early on, although H and He dominate the
composition of the outer ejecta. This is because the H content should be small
(only $\sim 0.05 \Msun$) and the \HeI\ levels giving rise to the strongest lines
are not excited early on. These levels are excited mostly via non-thermal
processes, but the fast particles that are responsible for the excitation are
produced deeper in the ejecta, and cannot penetrate from the deeper regions when
the densities are still too high. The delayed development of He lines has been
observed before in SNe\,Ib [\eg SN\,2005bf \citep{anup05}; SN\,2008D
\citep{sod08,maz08,tan09}.  Below 6500\,\kms, silicon and oxygen dominate the
composition, and Fe-group elements are also significant. 

This behaviour of the \HI\ and \HeI\ lines shows that if H and He are present in
the ejecta they will at some point manifest themselves in the spectra. Since in
SNe\,Ic He lines are never seen \citep[\eg][]{taub06}, and H lines are not seen
in most SNe\,Ib, these two subtipes must contain very little He and H,
respectively.

\subsection{Models for the nebular phase}

In order to assess the properties of the inner parts of the ejecta, and thus to
determine both the \Nifs\ mass (in addition to the light curve results) and the
ejected mass, it is essential that late-phase spectra are modelled as well as
early-time ones. At advanced epochs, the SN ejecta become optically thin and
behave like a nebula. The gas is heated by the deposition of the $\gamma$-rays
and positrons that are emitted in the decay of \Cofs\ to \Fefs. Collisions of
the fast particles that are produced in the thermalization process excite the
gas, which is cooled via mostly forbidden-line emission. The gas conditions are
modelled with a non-local thermodynamic equilibrium (NLTE) code
\citep{axe80,rll,maz01}.

The late-time spectra of SNe\,Ib/c are dominated by a strong \OI\ 6300,
6363\,\AA\ emission line, the shape of which can be used to infer the geometry
of the ejecta and our orientation, thanks to the transparency of the nebula
\citep{maz05}. 

We modelled the two latest spectra of SN\,2003bg, obtained on 2003 Dec 16 and
23, corresponding to fiducial epochs of 293 and 300 days, respectively.  The two
spectra are rather similar. Besides \OI\ 6300, 6363\,\AA, other strong lines are
\MgI] 4570\AA, several [\FeII] lines which testify to the brightness of the SN,
and \CaII\ lines. 

The shape of the emission lines does not suggest the presence of strong
asymmetries, and the spectra can be modelled assuming a single density and
homogeneous abundances below a velocity of 5000\,\kms, as indicated by the width
of the strongest emission lines ([\OI], \MgI]).  Models for the two spectra
yield reasonably consistent results (Fig. 3). Given the somewhat uncertain flux
calibration we take the mean of the results of the two spectral fits.  The
spectra can be reproduced assuming an emitting mass of $\approx 1.6 \Msun$
located below a velocity of 5000\,\kms. The main constituent of the emitting
nebula is oxygen ($\sim 0.9 \Msun$). Other elements are C ($\approx 0.1 \Msun$),
Si ($\approx 0.3 \Msun$), S ($\approx 0.1 \Msun$), Ca ($\approx 0.02 \Msun$), Mg
($\approx 0.005 \Msun$), and \Nifs.  The mass of \Nifs\ in the nebula is
$\approx 0.16 \Msun$, as required by the simultaneous fit of the [\FeII] and
other lines. By the time the nebular spectra were observed most \Nifs\ had
decayed to \Cofs\ and \Fefs. 

The explosion model we used for the early-time models has a mass of $1.3 \Msun$
below 5000\,\kms. The result from the nebular spectra is slightly larger, 
indicating the presence of more mass at low velocities than included in a
spherically symmetric explosion model. This is often seen in SNe\,Ib/c [\eg
SN\,1994I \citep{sau06}; SN\,1998bw \citep{maz01,mae02}], and may be the result
of some asphericity in the explosion \citep{maenom03}, which is typical of HNe
and possibly of all core-collapse SNe \citep{leo06,mae08}. Indeed, a hint for
the presence of an inner core of material with higher density may be found in
the profile of the [\OI] 6300, 6363\,\AA\ line, which may be composed of a main
profile characterized by narrowly separated double peaks plus a central core and
is reminiscent of that of SN\,2002ap \citep{maz07}. A model including a density
discontinuity to reproduce this profile in detail may have a slightly smaller
total mass. Still, the fact that the deviation from parabolic profiles is very
small in all lines suggests that any asphericity is small and probably affects
only the innermost ejecta. 

The strength of the semi-forbidden \CaII] 7200\,\AA\ line with respect to the
allowed \CaII\ IR triplet suggests that the density was low. In our models, the
electron density is $n_e \sim 2\:10^6$\,cm$^{-3}$.  A weak, broad [\OI]
5557\,\AA\ line may be present in the data, and is only partially reproduced in
the synthetic spectra, which also suggests a slightly higher density. A weak,
broad emission to the red of [\OI] 6300, 6363\AA\ may be identified as \Ha, as
in SN\,1993J \citep{patat95}.

\subsection{Results of spectral modelling}

Combining the early- and late-time modelling results we find that a model of a
highly energetic explosion which ejected $\sim 4-5 \Msun$ of material with \KE
$\approx 5\:10^{51}$\,erg can reproduce the observations. Only a small mass of H
is present, $\sim 0.05 \Msun$, located at $v > 9000$\,\kms. A massive He shell
is located at $v > 6500$\,\kms, comprising $\sim 1-2 \Msun$.  The ejected CO
core mass is $\sim 3 \Msun$, of which $\sim 1.3 \Msun$ is oxygen. The rest is
mostly carbon ($\sim 0.15 \Msun$), neon ($\sim 0.5 \Msun$), silicon ($\sim 0.6
\Msun$), and sulphur ($\sim 0.2 \Msun$). The mass of \Nifs\ is $\sim 0.17
\Msun$, most of which is located inside of 5000\,\kms. The confinement of \Nifs\
to the lowest velocities explains the late excitation of the He lines. 

Layers of H and He at the highest ejecta velocities were found by \citet{bra02}
in most SNe\,Ib. In SN\,2003bg and in other SNe\,Ib the H mass must be larger,
so that it gives rise to strong spectral lines. Also, in the case of SN\,2003bg,
we find that H extends deeper than in most SNe\,Ib, similar to the type IIb
SN\,1993J \citep{patat95}. The inner distribution of helium is at the lower
limit of the \citet{bra02} sample. 

Early on, neither H nor He lines are clearly detected. An elegant explanation
of this behaviour of H was given by \citet{bra02}. Additionally, in the case of
SN\,2003bg, the early identification of H lines is rendered difficult by the
breadth of the spectral features. He lines only develop later, as non-thermal
excitation increases with time \citep[see][]{m&l98}. 

The velocity evolution of SN\,2003bg (Figure 4) is similar to that of other
SNe\,Ib/c, in particular the HN SN\,2002ap, although our estimate of the
explosion epoch is uncertain. It is also similar to SN\,2008D early on. The
velocities we derived are based on modelling the spectra, and are therefore
influenced by trying to reproduce both the overall spectral distribution and the
observed line width. Therefore, they are sensitive to the assumed epoch,
especially at the earliest phases.  This is an indirect way of assessing the
reliability of our choice of epoch. If the date of explosion was chosen to be
earlier than assumed here, the velocity required to fit the spectra would most
likely be smaller, and the synthetic lines would be less blended. A later epoch
of explosion, on the other hand, would result in even larger velocities,
especially at the earliest times, and the blueshift of the lines would be too
large.

\section{Light Curve Model}

We synthesized the bolometric light curve (LC) of the ejecta models as
constrained by spectral fitting, and compare them to the observed LC, which is
described in the accompanying paper \citep{ham09}.  We used the 1-D SN LC code
that was originally developed by Iwamoto et al. (2000). The code solves the
energy and momentum equations of the radiation plus gas in the co-moving frame,
and is accurate to first order in $v/c$. Electron densities and the electron
scattering opacity are determined from the Saha-Boltzmann equation. The energy
deposition from radioactive decays of the newly synthesized $^{56}$Ni and its
daughter $^{56}$Co is calculated with a gray $\gamma$-ray transfer code,
assuming an absorptive opacity of $0.05 Y_e$ cm$^{2}$ g$^{-1}$ (Swartz,
Sutherland \& Harkness 1995). Unlike the sophisticated but time-consuming
recipes in the original code, we adopted the approximation proposed by
G\'{o}mez-Gomar \& Isern (1996) for the Eddington factors, and fitted the TOPS
opacities (Magee et al. 1995) to find an empirical relationship between the
Rosseland mean and the electron scattering opacity. Such simplifications were
also made by the authors when modeling the LCs of other Type Ic SNe.

The best-fitting model LC shown in Fig. 5 ({\em solid line}) is characterized by
an ejected mass of $\sim 4.8 \Msun$, with a kinetic energy of $\approx
5\,10^{51}$\,erg and including $0.15 \Msun$ of \Nifs, whose distribution in the
ejecta was adjusted by hand. The density structure above 5000\,\kms\ was
obtained scaling model CO100 as constrained by our photospheric-epoch spectral
models. Below 5000\,\kms, the density was enhanced to the level required by the
nebular spectrum modelling. In order to retain the calculation accuracy, the
ejecta material with $v/c>0.1$ was cut out. This amounts to a negligible mass
because of the extremely low density at such high velocities. Except for
$^{56}$Ni, we adopted the abundance distribution of other elements derived from
the spectrum synthesis. In particular, H exists only at $v \gtrsim 9000$\,\kms.
The total H mass is only $\approx 0.05\,\Msun$, which is too small to affect the LC
shape by H recombination, = as expected. The model LC shown has a \Nifs\
distribution of $\approx 0.01 \Msun$ (or $\approx 1\%$ in mass fraction) between
10,000 and 20,000\,\kms, $\approx 0.04 \Msun$ ($\approx 2.5\%$) between 5,000 and
10,000\,\kms, $\approx 0.05 \Msun$ ($\approx 4.4\%$) between 3,000 and 5,000\,\kms,
and $\approx 0.05 \Msun$ ($\approx 4.4\%$) below 3,000\,\kms. Thus the total \Nifs\
mass derived from the light curve ($0.15 \Msun$ is similar to that obtained from
the spectral fits, but the distribution is different. The light curve model
requires more \Nifs\ at high velocities. 

For comparison, two other model LCs with different model parameters are also
shown in Fig. 5. The {\em dashed line} shows a LC obtained with the same density
structure and hence the same mass and kinetic energy as the best-fit LC, but
with the \Nifs\ distribution derived from the spectrum modelling. The total
$^{56}$Ni mass is then $\approx 0.18 \Msun$, of which $\approx 0.16 \Msun$ are
located below 5000\,\kms. This low-velocity \Nifs\ must be responsible for the
high peak and post-peak flux, which is above the early-time photometry by $\sim
0.3$\,mag, although the LC matches the very late photometry, as it should by
design, since the inner distribution of \Nifs\ was obtained from fitting the
nebular spectra. Moreover, there are only $\approx 0.002 \Msun$ of $^{56}$Ni above
10,000\,\kms, not enough to power a very fast early LC rise. 

The {\em dotted line} is the synthesized LC of a model without the density
enhancement below 5000\,\kms\ and with the \Nifs\ distribution constrained by
spectral modeling. The total ejecta mass, kinetic energy, and \Nifs\ mass are
$\approx 3.9 \Msun$, $\approx 5\, 10^{51}$\,erg, and $\approx 0.12 \Msun$,
respectively. Although the energy available is smaller, the LC peak is actually
as as bright as in the other models because less ejecta material is heated.
However, because of the small ejecta mass relative to the kinetic energy, the LC
peak is too narrow and the LC drops too rapidly when compared to the
observations. As a further test of our model, we plot in the {\em inset} of Fig.
5 the evolution of the photospheric velocity of the three models as estimated
approximately by the LC code. They all reproduce well the values that were
determined by spectral modeling ({\em stars}).

While we cannot exclude the possibility that other \Nifs\ distributions, with
some tuning, may also give a good LC fit, it would be a futile attempt to try to
pin down the $^{56}$Ni distribution so precisely in a one-dimensional model.
Nonetheless, the distribution we adopted results in a rapid LC rise, and in a
maximum luminosity and a broad LC peak that match the observations well. After
about day 150 the model LC lies below the observations by $\sim 0.2$\,mag, but
the slope is comparable.

\section{Discussion}

The properties of SN\,2003bg derived from our analysis suggest that the SN was
the explosion of a massive star, clearly more massive than the Type IIb
SN\,1993J. The large explosion energy (\KE\,$\approx 5\:10^{51}$\,erg) qualifies
SN\,2003bg as a hypernova, in line with the broad lines detected in the earliest
spectra. If the ejected mass was $\sim 4-5 \Msun$, of which $\sim 1-2 \Msun$
were He and only $\sim 0.05 \Msun$ was H, the star had lost only its H envelope.
Assuming that the compact object formed at the time of core-collapse had a mass
between $1.4 \Msun$ (if it was a neutron star) and $2 \Msun$ (if it was a black
hole), we have a He star of $\sim 5.5-7 \Msun$ at explosion. This would imply a
CO core of $\sim 4-5 \Msun$ and a star of main-sequence mass $M_{ZAMS} \sim
22-25 \Msun$, very similar to the star that exploded as SN\,2002ap \citep{maz06}
and also similar to, or perhaps somewhat less massive than the progenitor of the
XRF-SN\,2008D \citep{maz08,tan09}. SN\,2003bg had a similar explosion energy as
these two other SNe, but because of the larger \Mej\ very high velocities were
not reached, and spectral lines were not as broad. This situation (low
velocities, larger overlying mass) also would have made the generation of a hard
X-ray or $\gamma$-ray transient very unlikely. 

The \Nifs\ distribution derived from the light curve model is different from
that obtained from the spectra. Most of the \Nifs\ obtained from the spectral
models is required to power the nebular spectra. If a higher abundance of \Nifs\
was used outside of the 5000\,\kms\ zone sampled by the nebular spectra, the
synthetic [\OI]\,6300\AA\ line would be too broad, unless \Nifs\ and oxygen were
physically separated. This would be a signature of asymmetry in the explosion,
as was seen in, \eg, SN\,1998bw \citep{maz01}.

SN\,2003bg falls on the relations between stellar main sequence mass and,
respectively, mass of \Nifs\ synthesised and \KE\ produced in the explosion
\citep{nom05}, as shown in Figures 6 and 7, suggesting that these relations hold
not only for SNe\,Ic but also for less stripped core-collapse SNe.

Given our estimate of the progenitor mass, the compact remnant is likely to have
been a black hole, but it also could have have been a neutron star.
Overenergetic explosions are expected also in this case if the neutron star is a
magnetar \citep{bucc08}.

However, the presence of significant CSM gave rise to strong radio emission in
SN\,2003bg \citep{sod06}. What is the difference between SN\,2003bg and
SN\,2006aj? SN\,2006aj had lost its He envelope, and possibly part of its CO
core. Thus the circumstellar material inside of which SN\,2006aj exploded was
composed mostly of oxygen \citep{maz06} SN\,2003bg only lost most of its H
envelope, and H was the main constituent of the CSM \citep{sod06}. Quite
possibly, SN\,2006aj was a binary star, which helped ejecting the envelope.
SN\,2003bg may have been a single star, and if it was an interacting binary the
effect of interaction was not as dramatic as in SN\,2006aj.

PAM acknowledges support from contracts ASI-INAF
I/023/05/0,  ASI I/088/06/0, and PRIN INAF 2006. 
JD acknowledges partial support by the National Natural Science Foundation of 
China (Grant No. 10673014) and by the National Basic Research Program of China 
(Grant No. 2009CB824800).
MH acknowledges support provided by NASA through Hubble Fellowship grant
HST-HF-01139.01-A (awarded by the Space Telescope Science Institute, which is
operated by the Association of Universities for Research in Astronomy, Inc., for
NASA, under contract NAS 5-26555), the Carnegie Postdoctoral Fellowship,
FONDECYT through grant 1060808, the Millennium Center for Supernova Science
through grant P06-045-F(funded by ``Programa Bicentenario de Ciencia y
Tecnolog\'ia de CONICYT''and ``Programa Iniciativa Cient\'ifica Milenio de
MIDEPLAN''), Centro de Astrof\'\i sica FONDAP 15010003, and Center of Excellence
in Astrophysics and Associated Technologies (PFB 06). 
This research has been supported in part World Premier International Research
Center Initiative (WPI Initiative), MEXT, Japan, by the Grant-in-Aid for
Scientific Research of the JSPS (18104003, 18540231, 20540226) and MEXT
(19047004, 20040004).


\newpage



\begin{deluxetable}{rrcrrr}
\tablewidth{0pt}
\tabletypesize{\scriptsize}
\tablenum{1}
\tablecaption{Parameters of the early-time synthetic spectra}
\tablehead{\colhead{Date} &
\colhead{SN epoch} &
\colhead{$\log L$} &
\colhead{$v_{ph}$} &
\colhead{$f$(\HeI)}   &
\colhead{$f$(\HI)}    \\
\colhead{~} &
\colhead{[days]\tablenotemark{a}} &
\colhead{[erg s$^{-1}$]} &
\colhead{km s$^{-1}$} &
\colhead{~} &
\colhead{~}   }
\startdata
28 Feb 2003 &  3 &  41.47  &  22000 & $10^3$ & $10^3$ \\
 2 Mar 2003 &  5 &  41.82  &  18000 & $10^3$ & $10^3$ \\
 4 Mar 2003 &  7 &  41.95  &  14100 & $10^3$ & $10^3$ \\
12 Mar 2003 & 15 &  42.31  &   8800 & $10^3$ & $10^3$ \\
31 Mar 2003 & 34 &  42.29  &   6000 & $10^6$ & $2 10^3$ \\
 4 Apr 2003 & 38 &  42.24  &   5600 & $10^7$ & $2 10^3$ \\
 9 Apr 2003 & 43 &  42.21  &   5100 & $10^9$ & $3 10^3$ \\
10 Apr 2003 & 44 &  42.20  &   5000 & $10^9$ & $3 10^3$ \\

\enddata
\tablenotetext{a}{The epoch is given from the putative date of explosion, 
25 Mar 2003.}
\end{deluxetable}




\clearpage
\begin{figure}
\epsscale{0.99}
\plotone{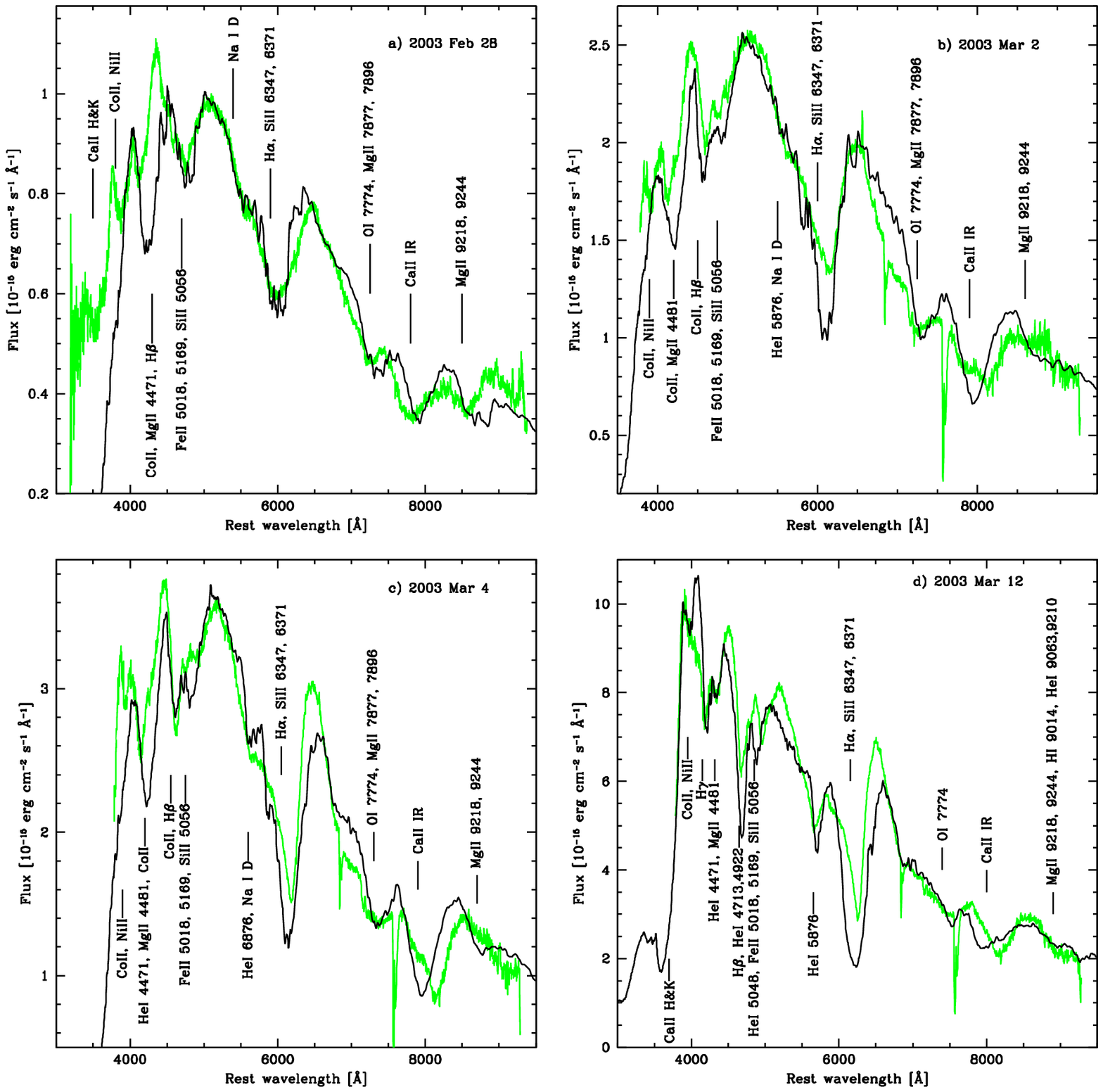}
\figcaption[f1.eps]{Observed spectra of SN 2003bg (green/black) compared to
synthetic models (black/grey) for pre-maximum epochs.
 \label{fig1}}
\end{figure}


\clearpage
\begin{figure}
\epsscale{0.99}
\plotone{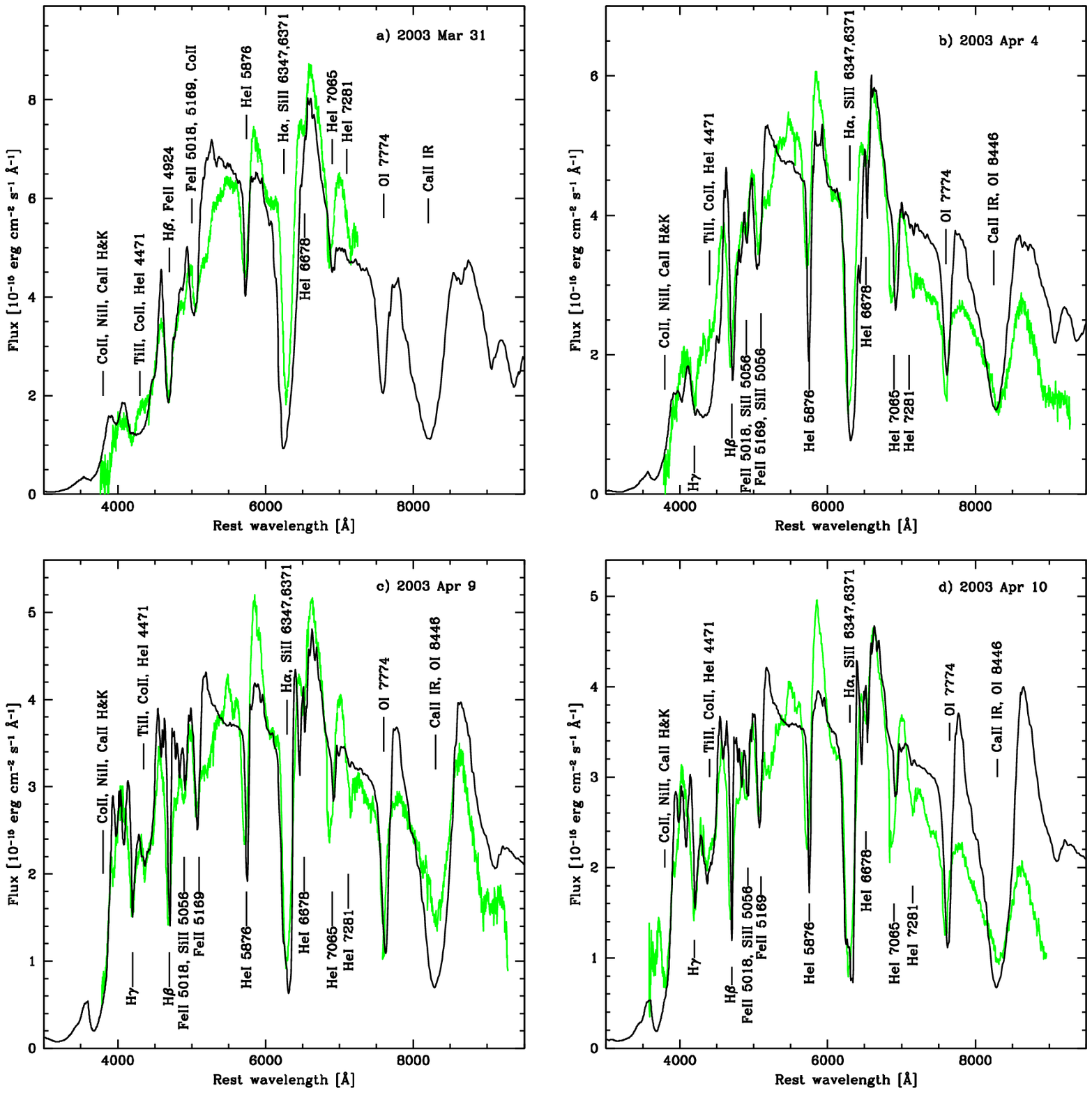}
\figcaption[f2.eps]{Observed spectra of SN 2003bg (green/black) compared to
synthetic models (black/grey) for post-maximum epochs. 
 \label{fig2}}
\end{figure}


\clearpage
\begin{figure}
\epsscale{0.99}
\plotone{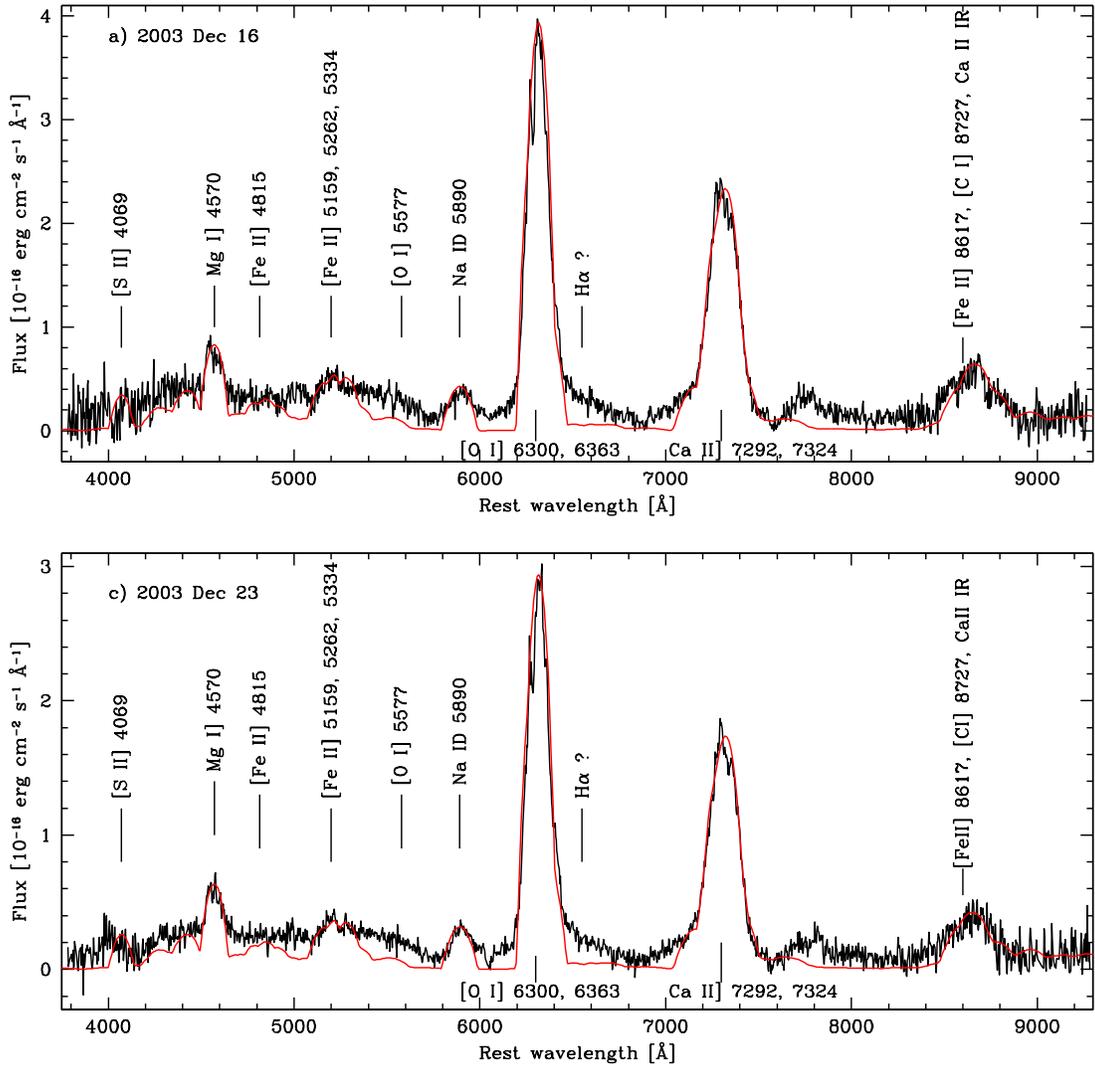}
\figcaption[f3.eps]{Observed nebular spectra of SN 2003bg (black) compared to
synthetic models (red/grey). 
 \label{fig3}}
\end{figure}


\clearpage
\begin{figure}
\epsscale{0.99}
\plotone{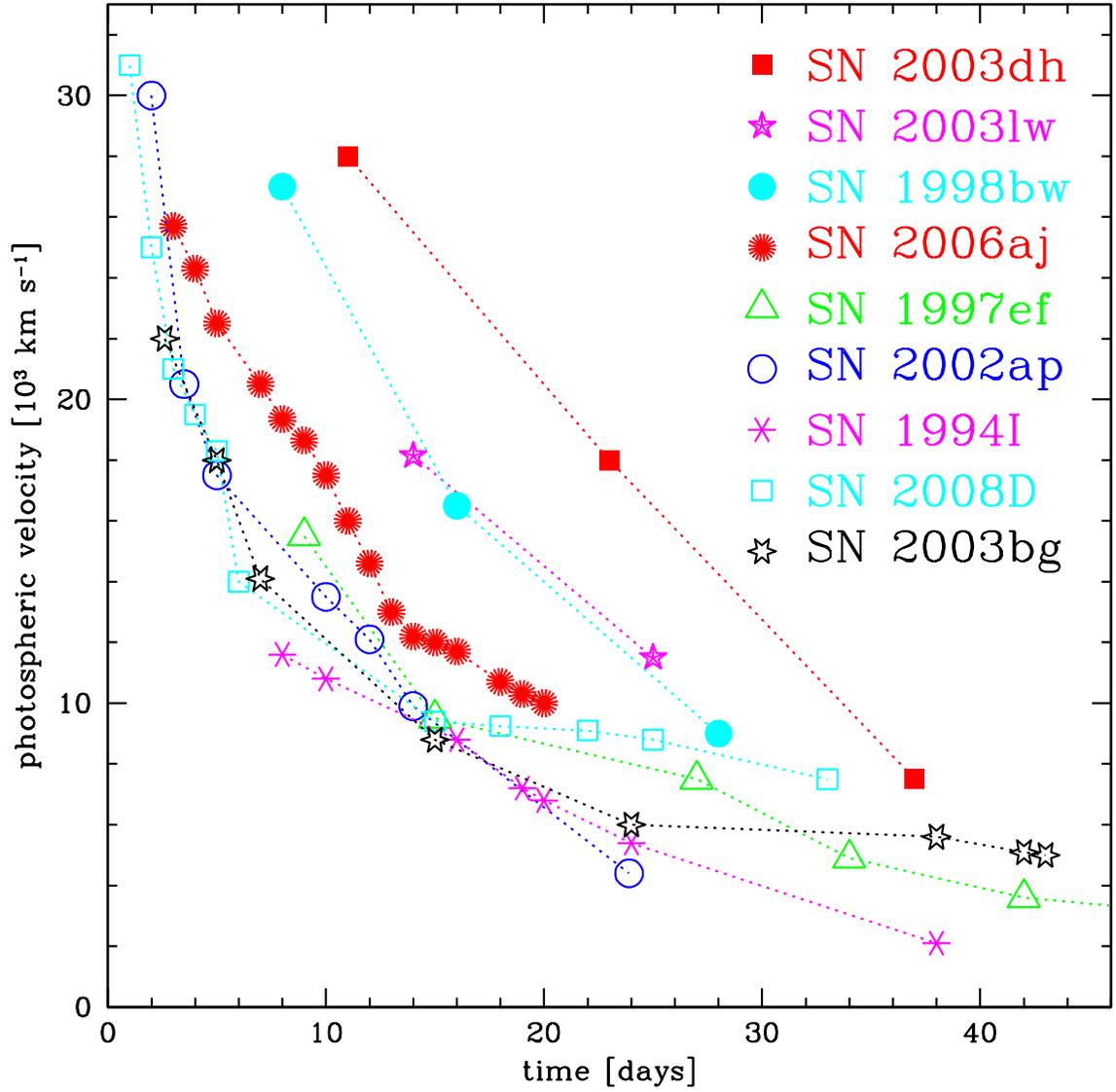}
\figcaption[f4.eps]{The evolution of the photospheric velocity obtained from
spectral models in a number of SNe Ib/c.
 \label{fig4}}
\end{figure}


\clearpage
\begin{figure}
\epsscale{0.99}
\plotone{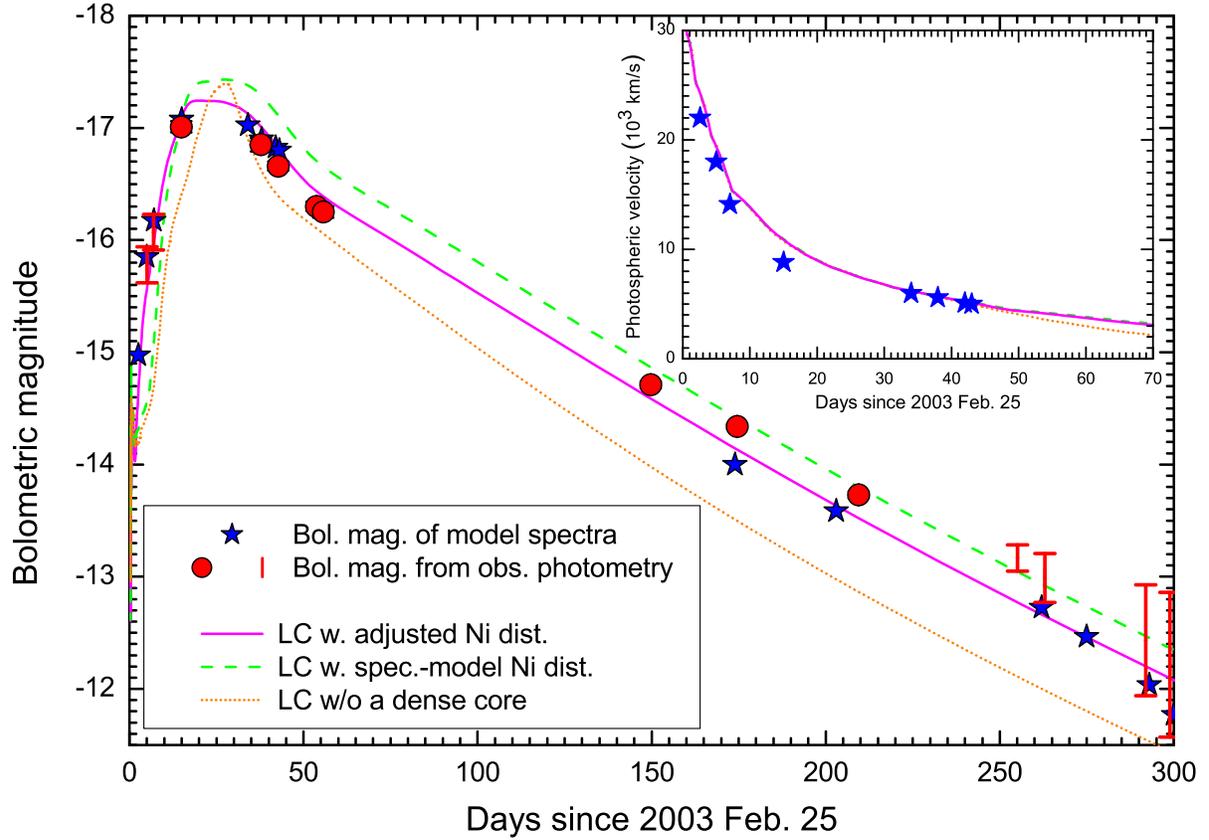}
\figcaption[f5.eps]{ The bolometric light curve of SN 2003bg compared to the
synthetic light curves of different models (see text). The inset shows the 
evolution of the photospheric velocity (as shown in Fig. 4) and the prediction 
of the various light curve models.
 \label{fig5}}
\end{figure}


\clearpage
\begin{figure}
\epsscale{0.99}
\plotone{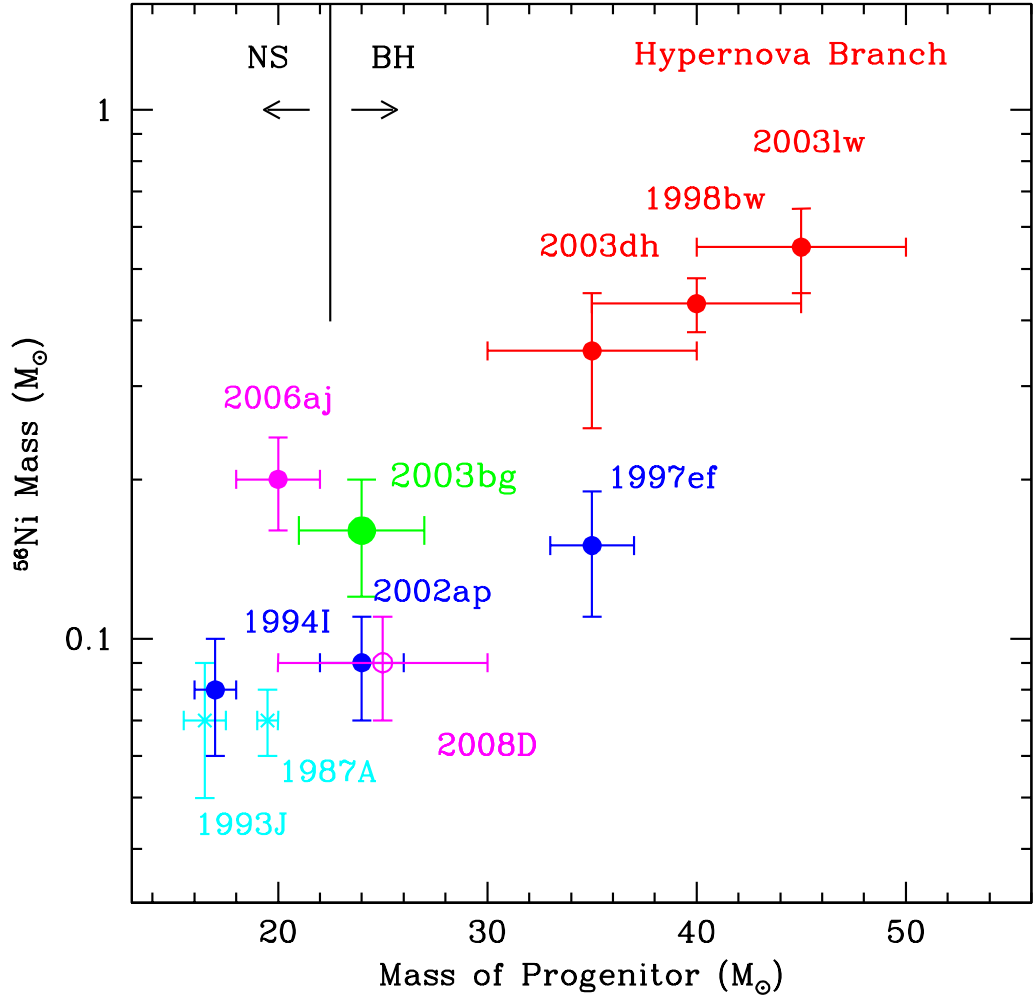}
\figcaption[f6.eps]{ The relation between progenitor mass and synthesised
 \Nifs\ mass in a number of Type Ib/c SNe. 
 \label{fig6}}
\end{figure}


\clearpage
\begin{figure}
\epsscale{0.99}
\plotone{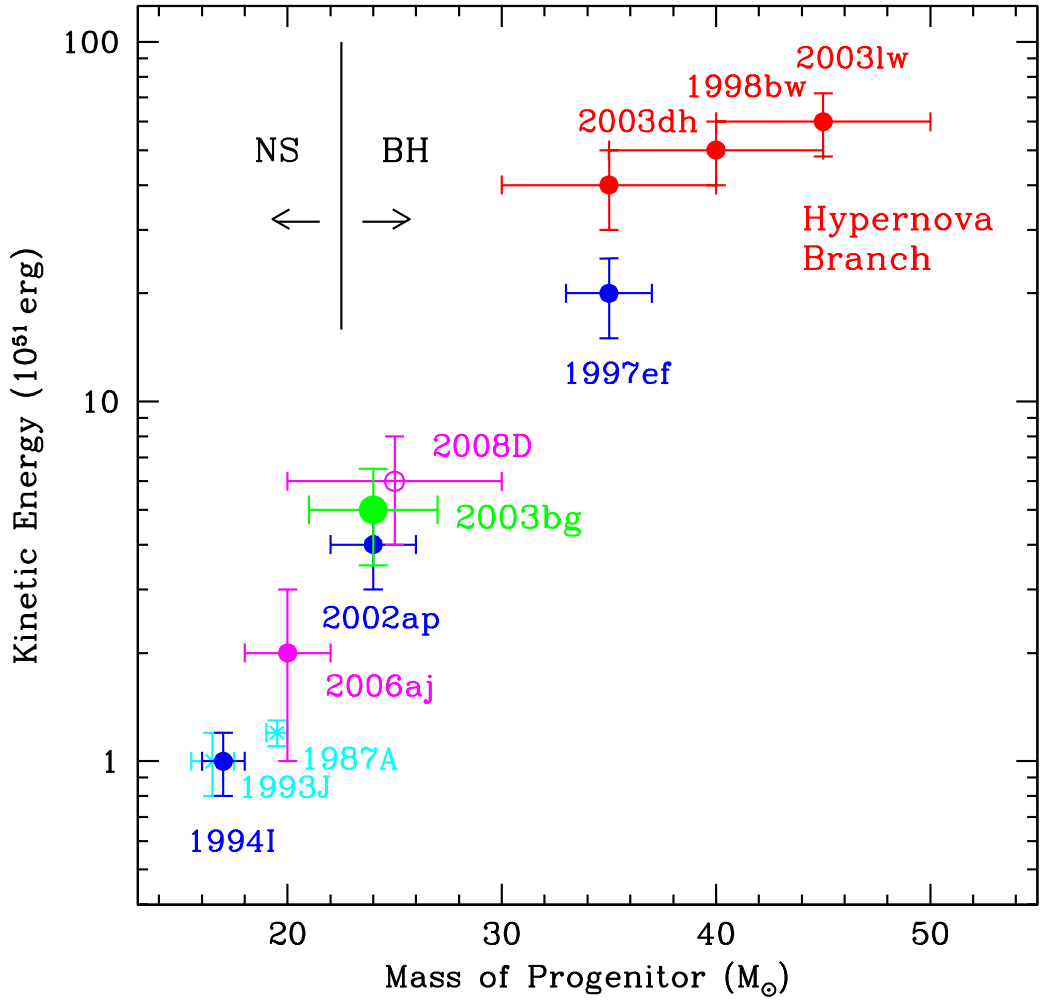}
\figcaption[f7.eps]{ The relation between progenitor mass and explosion kinetic
energy in a number of Type Ib/c SNe. 
 \label{fig7}}
\end{figure}

\end{document}